\documentclass{emulateapj}
\def\stacksymbols #1#2#3#4{\def\theguybelow{#2}
        \def\verticalposition{\lower#3pt}
        \def\spacingwithinsymbol{\baselineskip0pt\lineskip#4pt}
        \mathrel{\mathpalette\intermediary#1}}
\def\intermediary #1#2{\verticalposition\vbox{\spacingwithinsymbol
        \everycr={}\tabskip0pt
        \halign{$\mathsurround0pt#1\hfil##\hfil$\crcr#2\crcr
                \theguybelow\crcr}}}
\def\lta{\stacksymbols{<}{\sim}{2.5}{.2}}
\def\gta{\stacksymbols{>}{\sim}{3}{.5}}

\shorttitle{Star Formation in Local S0 Galaxies}
\shortauthors{Temi, Brighenti \& Mathews}

\begin{document}

\title{EVIDENCE OF STAR FORMATION IN LOCAL S0 GALAXIES: 
{\it Spitzer} OBSERVATIONS OF THE SAURON SAMPLE}

\author{Pasquale Temi\altaffilmark{1,2},
Fabrizio Brighenti\altaffilmark{3,4}, William
G. Mathews\altaffilmark{3} }
%\email{Pasquale.Temi@nasa.gov}
%\email{mathews@ucolick.org}
%\email{fabrizio.brighenti@unibo.it}

\altaffiltext{1}{Astrophysics Branch, NASA/Ames Research Center, MS
  245-6,
Moffett Field, CA 94035.}
\altaffiltext{2}{
Department of Physics and Astronomy, University of Western
Ontario,
London, ON N6A 3K7, Canada. ptemi@mail.arc.nasa.gov}
\altaffiltext{3}{University of California Observatories/Lick
  Observatory,
Board of Studies in Astronomy and Astrophysics,
University of California, Santa Cruz, CA 95064 
mathews@ucolick.org}
\altaffiltext{4}{Dipartimento di Astronomia,
Universit\`a di Bologna, via Ranzani 1, Bologna 40127, Italy 
fabrizio.brighenti@unibo.it}

\begin{abstract}
We discuss infrared {\it Spitzer} observations of early type galaxies 
in the SAURON sample at 24, 60 and 170$\mu$m. 
When compared with 2MASS $Ks$ band luminosities, 
lenticular (S0) galaxies exhibit a much wider 
range of mid to far-infrared luminosities then elliptical (E) galaxies. 
Mid and far-infrared emission from E galaxies is a combination 
of circumstellar or interstellar 
emission from local mass-losing red giant stars, 
dust buoyantly transported from the galactic cores into distant 
hot interstellar gas and dust accreted from the environment.
The source of mid and far-IR emission in S0 galaxies 
is quite different and is consistent with low levels of star
formation, 
0.02 - 0.2 $M_{\odot}$ yr$^{-1}$, in cold, dusty gaseous disks. 
The infrared 24$\mu$m--70$\mu$m color is systematically 
lower for (mostly S0) galaxies with known molecular disks. 
Our observations support the conjecture that 
cold dusty gas in some S0 galaxies is created by stellar mass loss 
at approximately the same rate that it is consumed by star formation, 
so the mass depletion of these disks by star formation will be slow.
Unlike E galaxies, 
the infrared luminosities of S0 galaxies correlate with both 
the mass of molecular gas and 
the stellar H$\beta$ spectral index,
and all are related to the recent star formation rate. 
However, star formation rates estimated from the H$\beta$ 
emission line luminosities $L_{H\beta}$
in SAURON S0 galaxies are generally much smaller.
Since $L_{H\beta}$
does not correlate with 24$\mu$m emission from dust
heated by young stars, optical emission lines appear to be
a poor indicator of star formation rates in SAURON S0 galaxies. 
The absence of H$\beta$ emission may be due to a relative 
absence of OB stars in the initial mass function 
or to dust absorption of H$\beta$ emission lines.
\end{abstract}


\section{Introduction}

In the discussion below 
we extend our earlier analysis of 
{\it Spitzer} infrared observations of elliptical galaxies (E) 
(Temi, Brighenti \* Mathews, 2007a)
to include lenticular galaxies (S0). 
The data set we consider here is further restricted to 31 
early type E (19) and S0 (12) galaxies that have been observed 
by both {\it Spitzer} and SAURON. 
The SAURON sample 
is generally representative and 
has received much attention from observers
at all wavelengths, allowing detailed comparisons of 
{\it Spitzer} infrared observations with many
different galactic attributes: stellar populations, 
nature and extent of cold gas, dust content, etc.

Our principal objective is to explore 
evidence of recent star formation 
in local S0 galaxies at mid and far infrared wavelengths 
and to contrast this with the apparent lack of measurable 
current star formation in the sample E galaxies.
Many S0 galaxies in the {\it Spitzer}-SAURON
sample have attributes consistent with recent star formation: 
significantly larger mid and far infrared luminosities, 
large masses of molecular gas, and 
younger stellar ages as inferred 
from optical absorption line spectra.
{\it Spitzer}-SAURON 
S0 galaxies having the largest far-infrared luminosities 
are known to contain rotationally supported disks of 
molecular hydrogen with masses $\gta 10^8$ $M_{\odot}$.
Using the 24$\mu$ emission as a guide, 
stars are forming in many sample lenticular galaxies  
at very modest rates, 0.02 - 0.2 $M_{\odot}$ yr$^{-1}$. 
For some of these S0 galaxies this level of star formation 
is comparable to the rate that gas is globally supplied from 
an old population of evolving stars, 
suggesting a quasi-steady state.
In their optical spectra 
our sample S0 galaxiess exhibit Balmer absorption lines from 
young A-F stars, 
but the Balmer line emission 
from ionized gas associated with more massive O-B stars 
can be lower than expected. 
In addition, when star formation rates are estimated 
using standard empirical procedures, the rates determined 
from H$\beta$ emission line luminosities in {\it Spitzer}-SAURON
S0 galaxies can be much less than those derived from 24$\mu$m 
luminosity.
The cold, dusty molecular disks in some S0 galaxies may 
originate from stellar mass loss, 
may be a vestige of a more substantial disk during a previous
incarnation as a more normal spiral galaxy,
or may have been acquired by a merger event.

\section{The SAURON-Spitzer Sample}

In Table 1 we list 34 (non-dwarf) 
E and S0 galaxies in the SAURON sample that are 
also in the public Spitzer archive.
This table contains additional details 
about each galaxy that will be discussed below. 
The SAURON sample is not complete 
but was chosen to include nearby 
early type galaxies representing a limited range of 
absolute magnitude, ellipticity, environments, etc. 
as described in detail by de Zeeuw et al. (2002). 
The {\it Spitzer}-SAURON sample that we discuss here 
includes additional imponderable attributes imposed by 
infrared observers who contributed to the {\it Spitzer} archive.

\section{Observations and Reduction}

Far infrared data presented here were obtained with the
Multiband Imager Photometer (MIPS) (Rieke et al. 2004) on
board the {\it Spitzer} Space Telescope (Werner et al. 2004)
in three wavebands centered at 24, 70 and 160 $\mu$m.
These data, collected from the {\it Spitzer} public
archive, do not form an homogeneous and uniform data set
in terms of image depth and observing mode.
Table 1 lists the program ID and PI name of the original
{\it Spitzer} observing program from which sample galaxies 
have been selected. The reader is referred to these programs 
to obtain details on the observing modes and imaging strategy,
as well as the on-source integration time for each target in 
the sample. As an example, Virgo Cluster galaxies recorded under
the guaranteed time program (PID 69, PI G. Fazio) reach
a relatively low sensitivity of 0.5MJy/sr and 1.1MJy/sr
(1$\sigma$) at 70 and 160 $\mu$m, respectively, while other 
galaxies (i.e. PID 20171, PI P. Temi) have deeper maps at
a sensitivity level of only 0.12MJy/sr and 0.3MJy/sr for
the same two wavebands. The SINGS data are recorded
in MIPS scan mode, covering a very large sky area
($30^\prime \times 10^\prime$ ), incorporating two separated passes 
at each source location. SINGS images correspond to maps with
intermediate sensitivity. Apart
from the SINGS observations, data have been acquired
in MIPS photometry mode, allowing appropriate coverage 
of the sources and their extended emission. 

Reduction of {\it Spitzer}-SAURON data follows the same procedure used
in Temi, Brighenti \& Mathews (2007a). Here we briefly summarize 
the basic processes involved in the reduction.
We started with the 
Basic Calibrated Data (BCD) products from the Spitzer 
Science pipeline (version 16.1) to construct mosaic 
images for all objects. 
Final calibrated images have been produced using the
MOPEX (Mosaicking and Point-source Extraction) package 
developed at the Spitzer Science Center (Makovoz et al. 2006).
MOPEX includes all the functions and steps necessary to 
process BCD data into corrected images and co-add them into 
a mosaic.
The major MOPEX pipeline used was the {\it mosaic pipeline} which
consists of a number of individual modules to be run in sequence
to properly perform the reduction.
We refer the reader to the MOPEX webpage for a detailed description
of each module. Since the data sets presented here have been 
acquired in different observing modes, the modules chosen 
in build-up the reduction flow and their parameter setup have
been carefully selected to properly remove mode dependent artifacts
in the final mosaic (outlier detection and median filtering). 

If foreground stars and background galaxies were present
in the ﬁnal mosaiced images they were edited before flux 
extraction was performed. 
These were identified by eye and cross-checked using surveys at other
wavelengths (Digital Sky Survey and 2MASS).
Flux densities were extracted from apertures that cover the entire 
optical disk (R25). Sky subtraction was
performed by averaging values from multiple apertures
placed around the target, avoiding any overlap with the
faint extended emission from the galaxy.
Statistical uncertainties related to sky subtraction are 
usually less than 1\% but
can be appreciable (tens of percent) for faint sources.
Observed infrared flux densities for each galaxy are
listed in Table 2.
Columns 2-4 contain the total flux
$F_{\lambda}\Delta \lambda$ (mJy)
at 24, 70 and 160 $\mu$m
within each MIPS passband
with respective widths $\Delta\lambda =$ 5.3, 19 and 34 $\mu$m.

Systematics in the MIPS calibration result in fluxes uncertain at
the 10\% level at 24 $\mu$m and 20\% at 70 and 160 $\mu$m. The uncertainties
listed in Table 2 do not include the systematic uncertainties.
Aperture corrections for extended sources
were applied to the fluxes as described in the Spitzer Observer's
manual.
The corresponding MIPS luminosities are
$L_{\lambda} = F_{\lambda}\Delta\lambda \cdot 4 \pi D^2$ where
distances $D$ are from Table 1.

\section{Observations}

\subsection{Star Formation Rates from 24 Micron Emission}

Figure 1 shows the relationship between the 
2MASS $Ks$ band luminosities of galaxies in Table 1 and their 
MIPS passband luminosities at 24, 70 and 160 $\mu$m. 
The $Ks$-band luminosities of E galaxies correlate tightly 
with $L_{24}$, but scatter dominates the correlations 
at 70 and 160 $\mu$m. 
These results are similar to those discussed by 
Temi, Brighenti \& Mathews (2007a) for a larger sample of 
E galaxies. 
Since E galaxies typically have de Vaucouleurs 
surface brightness profiles 
at 24 $\mu$m 
(Temi, Brighenti \& Mathews 2008), 
we conclude that most of the 24 $\mu$m emission 
originates in circumstellar dust around old red giant stars.
However, Temi, Brighenti \& Mathews (2007a,b) argue that 
E galaxy emission at 70 and 160$\mu$m 
arises from colder, truly interstellar dust. 
But the stochastic variation that produces the scatter 
in the lower panels of Figure 1 
is thought to be related to stochastic AGN energy 
releases from the central black hole 
that heats dusty gas in galactic cores which is then 
buoyantly transported out 5-10 kpc into 
the hot gas atmospheres within these galaxies 
(Temi, Brighenti \& Mathews 2007a,b). 
Only for ellipticals with the lowest $L_{70}$ and $L_{160}$ 
can the emission be explained by  dusty mass loss 
from local stars. 
Rare E galaxies that are unusually luminous 
at 70 and 160$\mu$m probably have experienced recent mergers, 
but, as discussed below, 
there is no strong evidence for recent 
mergers in the SAURON sample E galaxies.
Note that the vertical scatter of S0 infrared luminosities 
in Figure 1 significantly exceeds that of the E galaxies 
at all wavelengths. 
This is one of the strikingly different properties of S0 galaxies 
that we emphasize here.

In Figure 2 the infrared luminosities are normalized by $L_{Ks}$ 
(in solar unit) and the correlations at 24$\mu$m for E galaxies becomes much flatter. 
A least square fit to 24$\mu$m emission from E galaxies in 
Figure 2  gives 
\begin{equation}
L_{24} = 10^{29.4 \pm 0.5}L_{Ks}^{1.01 \pm 0.04}~~~{\rm erg~s}^{-1}.
\end{equation}
Evidently the nearly constant 
Log$(L_{24}/L_{Ks}) = 29.4$ for E galaxies represents the standard 
infrared SED of old stellar populations and their associated 
circumstellar dust. 
We show below that the departure of individual E galaxies from 
this relation are uncorrelated with H$\beta$ emission line
luminosity. 
Therefore, it is unlikely that the small variations of 
either infrared or optical line emission 
in SAURON E galaxies are 
related to appreciable star formation or recent mergers with 
star-forming dwarf galaxies.

While a few S0 galaxies appear to have old stars 
with little or no current star formation, as in many E galaxies, 
others have significantly larger $L_{24}/L_{Ks}$ 
which we interpret as evidence of a subpopulation of younger stars. 
We imagine that interstellar dust in
cold (possibly rotationally supported) gas 
is heated by massive, hot stars 
that recently formed from this gas.
The large vertical scatter in the S0 galaxies 
at all three wavelengths in Figures 1 and 2 
suggests that the interstellar dust in these galaxies 
has a much different 
origin and environment than far-IR emitting dust in E galaxies.

In their study of star-forming galaxies 
Calzetti et al. (2007) propose a relation between 
the specific 24$\mu$m luminosity,
$(\lambda L_{\lambda})_{24} \equiv \lambda_{24}L_{\lambda_{24}}$,
and the star formation rate, 
\begin{equation}
SFR(M_{\odot}~\rm{yr}^{-1}) = 
1.24 \times 10^{-38}[(\lambda L_{\lambda})_{24}
~({\rm erg~s}^{-1})]^{0.8850}.
\end{equation}
that can be used to find the SFR from global observations of 
$(\lambda L_{\lambda})_{24}$. 
Values of $(\lambda L_{\lambda})_{24}$ for our sample galaxies 
are plotted against $L_{Ks}$ in Figure 3.

We shall assume that this SFR-$(\lambda L_{\lambda})_{24}$ 
relation from Calzetti et al. holds for the S0 galaxies 
in Figure 3, but a correction must be made for the relatively 
strong 24$\mu$m contribution 
from old stars in these early type lenticular galaxies.
Unlike the star-forming galaxies considered by 
Calzetti et al, where almost all of the 24$\mu$m emission
evidently comes from interstellar dust irradiated by young hot stars, 
much or most of the 24$\mu$m emission from E and S0 galaxies 
originates in circumstellar dust surrounding old, mass-losing 
red giant stars. 
To remove this background 24$\mu$m emission, 
we determine the SFR from the 
$(\lambda L_{\lambda})_{24}$ emission in excess of that from 
the old stellar populations in 
our sample E galaxies collectively assumed to have $SFR = 0$.
The correlation between the specific stellar 24$\mu$m 
luminosity for old E galaxies and $L_{Ks}$, 
\begin{equation}
{\rm Log}(\lambda L_{\lambda})_{24,corr}
=  (1.01 \pm 0.05) {\rm Log}L_{Ks} + 30.1 \pm 0.5,
\end{equation}
is shown as a dashed line in Figure 3.
Therefore to estimate the star formation rate in S0 galaxies 
that lie significantly above this correlation in Figure 3, 
we enter the excess luminosity
\begin{equation}
\Delta(\lambda L_{\lambda})_{24} = 
(\lambda L_{\lambda})_{24} - (\lambda L_{\lambda})_{24,corr}
\end{equation}
in the SFR-$(\lambda L_{\lambda})_{24}$ relation above.
The resulting SFRs for galaxies with positive 
$\Delta{\rm Log}(\lambda L_{\lambda})_{24}$ 
are listed in the last column of Table 2. 
If $SFR = 0$ is assumed for E galaxies, 
the mean scatter among the SFRs evaluate for elliptical galaxies in
the upper section of Table 2, $0.014$ $M_{odot}$ yr$^{-1}$, 
suggests that the star formation rates for S0 galaxies 
(in the bottom section) are uncertain to
about $\pm 0.02$ $M_{odot}$ yr$^{-1}$.
The SFRs for these local S0 galaxies 
are all less than 1 $M_{\odot}$ yr$^{-1}$.

If the 70 and 160$\mu$m emission from S0 galaxies is due to 
the same process that emits at 24$\mu$m, such as star formation, 
we would expect the 70 and 160$\mu$m luminosities to correlate 
with 24$\mu$m residuals 
$\Delta{\rm Log}(\lambda L_{\lambda})_{24}$. 
This correlation is shown in Figure 4. 
Evidently, 
the same young stars in S0 galaxies 
that heat nearby interstellar dust grains to 
emit at 24$\mu$m 
also heat more distant grains that emit at longer wavelengths. 
However, when the 24$\mu$m residuals for S0 galaxies 
are plotted against 70 and 160$\mu$m luminosities 
that are normalized with the $Ks$-band luminosities, 
as in Figure 5, the correlation is greatly improved.
This figure suggests that $L_{FIR}/L_{Ks}$ is likely to be 
an equally good measure of the star formation rate 
in S0 galaxies as the 24$\mu$m residuals.  
Since Log$(L_B/L_{Ks})$
(from Table 1) does not increase systematically with 24$\mu$m residuals
or SFR(24$\mu$m), we conclude that the low level of star formation
in S0 galaxies
produces little observable additional $B$-band (or $Ks$-band) emission.
Consequently, $L_{Ks}$ is a good measure of the mass of old
stars and is insensitive to the small SFR in these S0 galaxies.

But Figure 5 informs us about a deeper relationship between 
star formation, nuclear molecular disks, 
and the mass of old stars in S0 galaxies. 
Young, Bureau \& Cappellari (2008) have studied 
compact CO-emitting molecular disks 
in four SAURON S0 galaxies: 
NGC 3032, 4150, 4459 and 4526. 
These disks typically have kpc-sized radii, total masses 
$0.5 - 5 \times 10^8$ $M_{\odot}$ and 
mean surface densities 100 - 200 $M_{\odot}$ pc$^{-2}$.
In NGC 3032, the S0 with the largest 
apparent star formation rate in Figure 5, 
the molecular disk 
(disk radius: 1.5 kpc, mass: $5 \times 10^8$ $M_{\odot}$) 
is counterrotating and must have been 
acquired by a galaxy merger in the past. 
The smaller molecular disk in NGC 4150 
(disk radius: 0.5 kpc, mass: $0.55 \times 10^8$ $M_{\odot}$)
is kinematically oriented 
with a dusty disk that obscures optical light.
However, NGC 4150 contains a counter-rotating 
core of younger stars having a spatial 
orientation that differs curiously from that of the molecular gas 
and rotates in the opposite sense. 
This suggests that the molecular disk in NGC 4150 
also resulted from one or more recent mergers.

In the remaining two S0 galaxies observed by 
Young, Bureau \& Cappellari (2008) -- 
NGC 4459 
(disk radius: 0.67 kpc, mass: $1.6 \times 10^8$ $M_{\odot}$)
and NGC 4526
(disk radius: 1.1 kpc, mass: $5.7 \times 10^8$ $M_{\odot}$) -- 
the molecular disks are well-aligned with the rotating 
system of old stars.
These appear to be more normal S0 galaxies. 
NGC 4459 and 4526, together with 
NGC 3156 and 3489, form a tight group of four S0 galaxies 
in Figure 5 having nearly 
the same $L_{FIR}/L_{Ks}$ and 24$\mu$m residuals, but with SFRs 
greater than normal elliptical galaxies. 
It is remarkable that among these four S0 
galaxies both $L_{FIR}$ and $L_{Ks}$
vary by about an order of magnitude yet they share the 
same location in Figure 5.
Since the mass of old stars increases with $L_{Ks}$ 
and $L_{FIR}$ depends on the total mass of dust and 
star-forming cold gas, 
it is possible that the star-forming gas in these S0 galaxies 
is being resupplied by mass ejected from old evolving stars. 
To explore this possibility, as $L_{Ks}$ varies over the range 
$10^{10} - 10^{11}$ (Figure 3), 
the rate that mass is ejected from old stars varies by   
about ${\dot M}_* \approx 0.01 - 0.1$ $M_{\odot}$ yr$^{-1}$ 
where we assume $(M_*/L_{Ks}) = 0.60$ 
(Bell et al. 2003) 
for an old stellar population with a Kroupa-like IMF
and 
$dM_*/dt \approx 0.15 M_{*}/(10^{11} M_{\odot})$
${M_{\odot}}_*$ yr$^{-1}$ (Mathews 1989).
Within the uncertainties involved,
we see that the mass supply from old red giant stars in 
these four star-forming 
SAURON S0 galaxies 
(${\dot M}_* \approx 0.01 - 0.1$ $M_{\odot}$ yr$^{-1}$)
is nearly equal to the SFR estimated 
from the 24$\mu$m dust emission. 
Therefore it is possible that the internal production of 
gas in these four S0 galaxies maintains rotationally supported 
disks of cold dusty gas that are forming young stars 
at the same rate, keeping the total mass of cold gas and 
dust rather constant with time. 

The two S0 galaxies with the largest 
24$\mu$m residuals and $L_{FIR}/L_{Ks}$ in 
Figure 5 -- NGC 3032 and 4150 -- 
have evidently acquired cold gas 
and dust in recent mergers, 
and their current star formation rates 
exceed the rate that mass is supplied by stellar mass loss. 
As star formation proceeds in these galaxies, 
we speculate that they may evolve downward in Figure 5 
(NGC 3032 $\rightarrow$ NGC 4150 $\rightarrow$ NGC 3156) 
until they reach the slowly-evolving, 
metastable situation described above 
for the four galaxies including NGC 3156 
where the SFR nearly balances the rate of stellar mass loss. 
This decreasing trend of star formation in cold nuclear 
disks may extend further to the only 
two elliptical galaxies in our sample known to contain 
molecular gas -- NGC 2768 and 4278 -- which also have 
the largest $L_{FIR}/L_{Ks}$ among the E galaxies in Figure 5.
When the stellar mass loss becomes comparable to the 
SFR, progress along this downward trend in Figure 5 
is expected to become slower.
Finally, since the stellar mass loss rate in an old single stellar 
population decreases with time ($\propto t^{-1.3}$), 
we speculate that the oldest merger-free S0 galaxies 
are those in Figure 5 that have joined the cloud of 
elliptical galaxies in which the cold nuclear disks have no 
observable CO or far-IR emission.

As seen in the infrared color-color plot in Figure 6
the infrared color Log$(L_{24}/L_{70})$ 
differs for those galaxies in Figure 5 that are known 
to contain molecular gas. 
Since these 
mildly star-forming galaxies have the lowest Log$(L_{24}/L_{70})$ 
colors of the entire Spitzer-SAURON sample, this 
color could be used to identify other E and S0 galaxies that 
contain similar molecular disks.

\subsection{Correlations of 24$\mu$m Emission with H$\beta$ Index and
  Molecular Mass}

From their analysis of SAURON galaxies McDermid et al. (2006) 
showed that the mean stellar age of S0 galaxies 
is systematically much 
less than that of E galaxies. 
We now show that 
this result is consistent with our observations at MIPS wavelengths.
Young stellar ages correspond to larger H$\beta$ Lick 
(absorption) indices in optical spectra. 
In Figure 7 we show that the optical H$\beta$ index in S0 
galaxies (evaluated within $R_e/8$) 
correlates positively with bandpass luminosities 
at 24, 70 and 160 $\mu$m, 
suggesting that a substantial fraction of this infrared 
radiation comes from interstellar dust irradiated and 
heated by young stars.

The correlations in Figure 7 show that the contribution of 
young stars to the H$\beta$ index in S0 galaxies increases with 
$L_{FIR}/L_{Ks}$ since this latter ratio correlates 
with the 24$\mu$m residuals (Fig. 5) and the SFR.
We expect an even stronger correlation if 
H$\beta$ is plotted against the 24$\mu$m residuals. 
This expectation is confirmed in the upper panels of Figure 8 
for H$\beta$ indices determined within either $R_e/8$ or $R_e$ 
of the galactic centers. 
S0 galaxies with H$\beta \gta 2$ contain 
systematically  younger stars.
Further confirmation that 
$\Delta{\rm Log}(\lambda L_{\lambda})_{24}$
is related to the star formation rate 
is provided by the lower panel of Figure 8, 
showing that the mean stellar age from 
Table 1 decreases with increasing residuals in Figure 3.
The three E galaxies in Figure 8 
with mean stellar ages less than 5 Gyrs -- NGC 2768 and 5831 -- 
are somewhat anomalous in not having larger residuals 
$\Delta{\rm Log}(\lambda L_{\lambda})_{24}$.

While observations of the HI gas mass in 
our {\it Spitzer}-SAURON sample are too sparse 
to consider here, more observations are available of their molecular
mass
(Combes et al. 2007). 
Figures 9 and 10 show that the molecular mass $M_{H_2}$ 
in S0 galaxies increases with all Spitzer infrared passband
luminosities 
$L_{FIR}$ and also with $L_{FIR}/L_{Ks}$, although the scatter 
in Figure 10 is rather large (see also Combes et al. 2007, who found a
similar correlation using IRAS data).
A spurious unit slope correlation in Figure 9 could result 
in errors in the distances to the galaxies 
and this may account for the degrated correlation in Figure 10.
Postive correlations in these Figures would be 
expected if the emitting dust 
in S0 galaxies is embedded 
inside molecular gas in which young stars are forming. 
The two elliptical galaxies containing molecular gas in these 
figures, NGC 2768 and 4278, have small positive residuals
in Figure 3, 
but are not 
otherwise exceptional.

\subsection{Disparity between SFR from 24$\mu$m 
and H$\beta$ Emission Luminosities} 

One of the many advantages of 
SAURON integral field spectroscopy are improved measurements 
of the total luminosity of optical emission lines. 
In Figure 11 we plot the total luminosity in H$\beta$ emission 
against far infrared luminosity in each MIPS passband 
normalized with the $Ks$ band luminosity. 
E and S0 galaxies appear to behave in a different fashion 
in Figure 11. 
In our {\it Spitzer}-based sample of E galaxies  
we found positive correlations 
between the observed H$\beta$ emission line 
flux and the bandpass flux $F_{FIR}$  
at all three wavelengths, 24, 70 and 160$\mu$m 
(Temi, Brighenti \& Mathews 2007). 
However, when the FIR luminosity is normalized with $L_{Ks}$, 
as in Figure 11, the correlation between the H$\beta$ emission 
luminosity and $L_{24}/L_{Ks}$ disappears for the E galaxies. 
This indicates that the $Ks - 24\mu$m color is part of the 
normal SED of an old stellar population and that the 
H$\beta$ emission is uncorrelated with this color. 
At 70 and 160$\mu$m, however, E galaxy $L_{H\beta}$ correlates 
with $L_{FIR}/L_{Ks}$.  
These correlations are consistent with the analysis of
Temi, Brighenti \& Mathews (2007a,b) in which both H$\beta$ and 
FIR emission arise from buoyantly transported 
cooling regions widely dispersed in the hot interstellar gas 
of E galaxies.

By contrast, the H$\beta$ emission from 
S0 galaxies in Figure 11 appears not to correlate 
with $L_{70}/L_{Ks}$ or $L_{160}/L_{Ks}$ and exhibits larger overall 
$L_{FIR}/L_{Ks}$ than the E galaxies. 
Evidently H$\beta$ emission from S0 galaxies 
in the SAURON sample is generally unrelated to far infrared 
emission from dust. 
The absence of this correlation for S0 galaxies 
may seem surprising 
since massive young stars are the likely source of both 
optical line emission and dust heating with subsequent far-IR
emission,
as in the Milky Way disk. 
Perhaps some of the H$\beta$ emission is absorbed by dust.

In the top panel of Figure 12 we show this result in a different 
way by plotting the 24$\mu$m residuals in Figure 3 against the 
H$\beta$ emission line luminosity. 
The absence of any correlation for E galaxies indicates that 
there is no coordinated enhancement of emission from heated 
dust or optical emission lines that would 
be expected from star formation or from a recent merger with a 
star-forming galaxy containing gas 
and dust.
%\footnote{In their discussion of optical 
%line emission from SAURON 
%galaxies Sarzi et al. 2006 suggest that kinematical misalignments 
%between emission line and stellar velocities 
%in early type galaxies may arise from 
%recent accretion of interstellar gas, particularly in 
%more spherical, slowly rotating E galaxies. 
%However, quasi-random non-stellar emission line velocities 
%with amplitudes $\pm 150$ km s$^{-1}$ are often observed 
%in slit spectra of E galaxies (e.g. Caon et al. 2000) which have no 
%additional evidence of recent mergers. 
%These unusual emission line velocities could arise through the 
%interaction of warm gas with the (unknown) velocity field 
%of the hot, virialized interstellar gas.}
This result is consistent with the finding of 
Temi, Brighenti \& Mathews (2005) that E galaxies with 
mean stellar ages (from Balmer indices) less than about 5 Gyrs 
often show no measurable enhancement at 24$\mu$m.
For E galaxies 
the H$\beta$ emission is clearly unrelated to star formation. 
A low level of H$\beta$ emission 
is always expected in E galaxies 
from stellar mass loss into the hot interstellar gas 
that is ionized by hot post-AGB stars until it thermally merges 
into the hot interstellar phase. 
Larger, more stochastic contributions to optical line emission 
in E galaxies can result from 
cooling dusty hot gas that has been buoyantly driven out of 
E galaxy cores by low energy AGN activity 
(Temi, Brighenti, \& Mathews 2007a,b). 
(Mergers as a source of strong E galaxy dust emission at 70 and
160$\mu$m 
are also possible, particularly when there is additional morphological 
or dynamical evidence of a merger event.) 

As before, S0 galaxies in the top panel of 
Figure 12 show no obvious correlation 
between $L_{H\beta}$ and 24$\mu$m residuals from Figure 3.
The correlations of 
24$\mu$m residuals $\Delta(\lambda L_{\lambda})_{24}$,
and therefore also SFR(24$\mu$m), 
with both the stellar H$\beta$ index (Figure 8) 
and the mass of molecular gas (lower panel of Figure 12) 
confirm that 24$\mu$m emission is likely to be a good measure 
of star formation in our sample S0 galaxies. 
But the lack of a correlation between $L_{H\beta}$ and 24$\mu$m
residuals in SAURON S0 galaxies suggests that the standard
observational method of determining the SFR from 
optical emission lines
(e.g. Kennicutt 1998; Calzetti 2007) is inaccurate or 
inappropriate for these galaxies which have relatively low levels 
of star formation. 

According to Calzetti (2007),
the current star formation rate can be determined 
from H$\beta$ emission line luminosities using 
\begin{equation}
SFR(H\beta) = 1.51 \times 10^{-41} L_{H\beta}~~~
({\rm ergs~s}^{-1})
\end{equation}
where we assume $L_{H\alpha} = 2.86 L_{H\beta}$ 
since accurate $H_{\alpha}$ line luminosities cannot be
measured with SAURON. 
(While this Case B $H{\alpha}/H{\beta}$ ratio is not expected 
to hold in the presence of dust absorption, it is unlikely to 
be seriously in error for escaping line radiation.) 
The SFR(H$\beta$) in equation 5 is derived using 
a Kroupa-like IMF extending to 120 $M_{\odot}$. 
Figure 13 shows that SFR(H$\beta$) is much less than SFR(24$\mu$m) 
for many of our SAURON sample galaxies. 
For galaxies with the largest SFR(24$\mu$m) -- NGC 3032, 
4526 and 4459 -- SFR(H$\beta$) is an order of magnitude 
smaller than SFR(24$\mu$m).
These are also the three galaxies in Table 1 with the largest 
masses of molecular gas, $1.6 - 5 \times 10^8$ $M_{\odot}$. 

Calzetti et al. (2007) also suggest a means of determining 
the star formation rate in the presence of dust 
extinction that employs both 
Balmer line and 24$\mu$m luminosities, SFR(24$\mu$m,H$\beta$).
However, we find that SFR(24$\mu$m,H$\beta$) is very nearly 
identical to SFR(24$\mu$m), i.e. the optical line emission 
is too low to affect the SFR from 24$\mu$m alone. 

If star formation occurs in intermittent bursts separated 
by more than the main sequence lifetimes of OB stars, 
$\sim10^7$ yrs, the SFR from HII emission line luminosities 
might often appear to be less than that from 24$\mu$m radiation.
But it is very improbable that star formation in 
these massive ($\sim10^8$ $M_{\odot}$) 
molecular disks would occur in globally coherent bursts.

Alternatively, 
the relative absence of H$\beta$ emission could be understood 
if massive stars are not forming in these S0 galaxies; 
Calzetti (2007) notes that SFR($H\beta$) drops 
by $\sim$5 if the maximum stellar mass is decreased from 
120 to 30 $M_{\odot}$. 
However, the stellar initial mass function (IMF) may be 
radically altered in molecular disks in S0 galaxies. 
Typical surface densities for the disks observed by 
Young, Bureau \& Cappellari (2008), 
100 - 200 $M_{\odot}$ pc$^{-2}$, correspond to
about 0.02 - 0.04 gm cm$^{-2}$ which is less than 
the threshold density $\sim1$ gm cm$^{-2}$ required 
by Krumholz and McKee (2008) for massive star formation. 

The column depths perpendicular to these molecular disks 
correspond to hydrogen columns of 
about $10^{22}$ cm$^{-2}$ for which the optical depth 
to dust absorption of H$\beta$ is about ten. 
Therefore it is conceivable that SFR(H$\beta$) is 
underestimated simply because ionizing massive stars form 
close to the central disk plane where line emission cannot 
escape.
However, the 24$\mu$m residuals and therefore SFR(24$\mu$m) 
correlate with the stellar H$\beta$ index so at least 
some of the young stars with Balmer (absorption) lines must be visible. 
For this obscuration model to work it would be necessary 
for A and F stars to be dynamically excited away from the 
disk plane so their radiation can escape. 
Further insights about the missing optical line emission 
in S0 galaxies might be gained by observing emission lines further 
in the infrared, like for example Br$\gamma$ (at 2.16$\mu$m),
which are less absorbed by dust.

\section{Discussion}

The star formation rates in Table 2 found from 24$\mu$m emission  
using the SFR-$(\lambda L_{\lambda})_{24}$ 
relation derived by Calzetti et al. (2007) 
suggests that the SFR in SAURON elliptical galaxies is 
$\lta 0.02$ $M_{\odot}$ yr$^{-1}$ and may in fact be zero.
However, many S0 galaxies in the SAURON sample have 
SFRs that exceed 0.02 $M_{\odot}$ yr$^{-1}$ and extend 
up to almost 0.2 $M_{\odot}$ yr$^{-1}$ for NGC 3032.

Our results complement those of 
Young, Bendo \& Lucero (2008) who studied 
MIPS infrared emission from a sample 
of E and S0 galaxies known to be very rich in molecular gas 
(M(H$_2$) $\approx 1.5 - 64 \times 10^8$ $M_{\odot}$).
They found, as we do, evidence of enhanced 24$\mu$m 
emission from many S0 galaxies and showed that 
some S0s are measurably extended at this wavelength 
having sizes comparable to CO molecular images.
This suggests that dust emitting at 24$\mu$m is indeed associated 
with star formation in cold molecular disks. 
Three of the galaxies in the H$_2$-rich sample of 
Young, Bendo \& Lucero (2008) are also in the SAURON 
sample -- NGC 3032, 4459 and 4526 -- having 
M(H$_2$) = 2.6, 1.6 and 4.0 $\times 10^8$ $M_{\odot}$ 
respectively (Table 1).

Combes, Young \& Bureau (2007) have detected CO emission 
from about 28\% of E or S0 galaxies in the SAURON sample.
Of the ten galaxies with CO detections shown in 
our Table 1, the majority (8) are S0 galaxies. 
They also find that CO-rich E and S0 galaxies have 
stronger H$\beta$ spectral indices.
This is consistent with our finding that both 
M(H$_2$) and the H$\beta$ index correlate with 
non-stellar 24$\mu$ emission in S0 galaxies.

Finally, the referee of our paper suggested that we 
also consider an analysis of the mid and far-infrared data 
in our sample
using the new two-dimensional kinematical classification 
of early-type galaxies proposed by Emsellem et al. (2007) 
which is based on SAURON data.
This new classification scheme is based on a single parameter,
$\lambda_R \equiv \langle R|V|\rangle /
\langle R (V^2 + \sigma^2)^{1/2} \rangle$, 
where R is the distance from the galactic center in the 
image plane, $V$ is the stellar radial velocity, $\sigma$ is the 
stellar velocity dispersion, and the two-dimensional averages 
are taken within the effective radius $R_e$.
Galaxies with $\lambda_R > 0.1$ are designated fast rotators and 
those having $\lambda_R < 0.1$ are slow rotators.
Our {\it Spitzer}-SAURON sample contains 9 slow and 22 
fast rotators. 
We made alternative versions of Figures 1 - 12 using 
this new classification scheme and determined that 
all correlations and conclusions based on the figures 
presented here remain unchanged, 
although the E-like slow 
rotators are fewer in number and more sparsely distribued 
in the alternative figures.
Nevertheless, some plots are marginally different. 
For example in the $\lambda_R$-versions of Figures 1-3, 
the slow rotators exhibit less scatter in the vertical
direction, with fast rotators lying both above and below, 
but this may be a result of the smaller number of slow rotators
(9) compared to E galaxies (19) in the ususal classification.
Nevertheless, we do not include the alternative figures here 
because they contain no important new information.
Furthermore, we think it may be premature to abandon 
the familiar morphological early-type classification scheme 
in terms of E and S0 galaxies 
since the new $\lambda_R$ classification 
parameter is currently unavailable for the vast majority of 
early-type galaxies.

\section{Conclusions}

Perhaps the most important conclusion from 
these {\it Spitzer} observations of infrared 
emission from early type SAURON galaxies is that there are 
significant differences between S0 and E galaxies 
in the relative content of dust and cold gas 
and star formation rates.
Only a small subset of S0 galaxies appear to share with 
E galaxies the same 
correlations between the 24$\mu$m luminosity, 
the H$\beta$ index, and molecular mass. 
Results from Sloan surveys of ``early type'' galaxies that combine 
S0 and E galaxies 
(e.g. Kaviraj et al. 2007)
may therefore be misleading if applied only to E galaxies.
It is likely that the source of cold gas in E and S0 galaxies 
is systematically quite different.
Rotationally supported
cold gas in S0 galaxies may be a relic of
their previous incarnation as late type spirals.

The threshold for measurable star formation rates 
in the SAURON sample is about 0.02 $M_{\odot}$ yr$^{-1}$. 
All S0 galaxies (except NGC 474, 1023, 7457 and 4382) 
exceed this value and all 
E galaxies (except NGC 4486 and possibly NGC 2768)
have lower SFRs.
We have shown that the 24$\mu$m luminosity is a good 
measure of the SFR and that it correlates well with 
other related parameters such as 
the H$\beta$ index, the mean stellar age determined 
from SAURON observations and the mass of molecular gas.

Positive residuals from 
the E galaxy correlation between  
$\lambda L_{\lambda}$ at 24$\mu$m and $L_{Ks}$ 
represent the emission from interstellar dust alone 
and are a measure of the galactic star formation rate. 
It is remarkable that these residuals 
for S0 galaxies correlate so 
tightly with normalized far-IR luminosities 
$L_{60}/L_{Ks}$ and $L_{170}/L_{Ks}$.
This suggests that the amount of dust, cold gas and 
old stars is rather finely tuned. 
The estimated mass loss rates from 
old stars in S0 galaxies 
are very nearly the same as the star formation rates 
derived from interstellar 24$\mu$m emission. 
This supports the notion that the mass of cold, dusty gas 
in some SAURON S0 galaxies is in an approximate steady 
state -- gas is lost by star formation at nearly the 
same rate that new gas is ejected from the dominant 
old stellar population.

However, the H$\beta$ emission line luminosity 
from S0 galaxies appears to be very low and 
unrelated to the star formation rate. 
(This is also true for the global $B - Ks$ colors of 
S0 galaxies in our sample.) 
Star formation rates determined from the mid-IR 
luminosity $(\lambda L_{\lambda})_{24}$ are often about 
an order of magnitude larger than those found from 
the H$\beta$ emission line luminosity, 
using the standard SFR recipes. 
Perhaps star formation in 
cold gaseous disks in early type galaxies does not 
extend to massive OB stars because the disk column 
densities fall below the Krumholtz-McKee threshold 
of $\sim1$ gm cm$^{-2}$. 
An alternative interpretation is that OB stars 
do indeed form but 
are deeply embedded in dust that absorbs H$\beta$ 
line emission. 
If this is the explanation, 
radiation must still 
escape from slightly older A-F type stars since 
the star formation rate based on 24$\mu$m emission 
correlates with the Lick H$\beta$ index in the 
spectra of young $\sim$A stars.
These non-ionizing stars may be 
less concentrated toward the central plane of the 
dusty molecular disks, allowing their radiation to escape.

\acknowledgements
This work is based on observations made with the Spitzer Space
Telescope, which is operated by the Jet Propulsion Laboratory,
California Institute of Technology, under NASA contract 1407.
Support for this work was provided by NASA through Spitzer
Guest Observer grant RSA 1276023.
Studies of the evolution of hot gas in elliptical galaxies
at UC Santa Cruz are supported by a Spitzer Theory Grant 
and an NSF grant for which we are very grateful. 
Finally we thank the referee for thoughtful remarks.

\clearpage

\begin{deluxetable}{clccrrrccrrc}
\tabletypesize{\scriptsize}
%\rotate
%\tablecaption{MIPS Far-infrared photometry }
\tablecaption{}
\tablewidth{0pt}
\tablehead{
\colhead{Name} & PI/Program ID &
\colhead{Type\tablenotemark{a}} &
\colhead{T\tablenotemark{b}} &
\colhead{D\tablenotemark{c}} &
\colhead{Log $L_B$} &
\colhead{Log $L_{Ks}$\tablenotemark{d}} &
\colhead{Log $M_{HI}$\tablenotemark{e}} &
\colhead{Log $M_{H_2}$\tablenotemark{f}} &
\colhead{$H\beta$\tablenotemark{g}} &
\colhead{Age}\tablenotemark{h} &
\colhead{Log $F_{H\beta}$\tablenotemark{i}}\\
\colhead{(NGC)} & & \colhead{} & \colhead{} &
\colhead{Mpc} & \colhead{($L_{B,\odot}$)} &\colhead{($L_{Ks,\odot}$)}
&\colhead{($M_\odot$)} &\colhead{($M_\odot$)} &\colhead{(\AA)}
&\colhead{(Gyr)} &\colhead{($erg\; s^{-1}\; cm^2)$}
}
\startdata

     0821 & Fabbiano/20371 & E6?      & -4.8 &  24.09 &  10.28 &  10.95 &\nodata &
     \nodata  &  1.53 & \nodata& \nodata \\
     2768 & Temi/20171 & E6:      & -4.3 &  22.38 &  10.63 &  11.25 &  8.23  &
     7.65  &  1.70 &  2.5   &  -13.55 \\
     2974 & Kaneda/3619 & E4       & -4.7 &  21.48 &  10.26 &  11.51 &\nodata &$ <
     7.59 $&  1.64 &  7.9   &  -13.35 \\
     3377 & Fazio/69 & E5-6     & -4.8 &  11.22 &   9.82 &  10.47 &\nodata &$ <
     6.88 $&  1.85 & \nodata&  -13.63 \\
     3379 & Fazio/69 & E1       & -4.8 &  10.57 &  10.11 &  10.89 &\nodata &$ <
     6.69 $&  1.43 & 11.5   &  -14.41 \\
     3608 & Fazio/30318 & E2       & -4.8 &  22.91 &  10.24 &  10.83 &\nodata &$ <
     7.51 $&  1.49 &  8.9   &  -14.77 \\
     4278 & Fazio/69 & E1-2     & -4.8 &  16.07 &  10.23 &  10.88 &  8.84  &
     7.50  &  1.23 & \nodata&  -12.86 \\
     4374 & Rieke/82 & E1       & -4.2 &  18.37 &  10.69 &  11.38 &\nodata &$ <
     7.15 $&  1.40 & \nodata&  -13.66 \\
     4458 & Fazio/30318 & E0-1     & -4.8 &  17.22 &  9.57  &  10.10 &\nodata &$ <
     7.28$&  1.63 & \nodata& \nodata \\
     4473 & Cote'/3649 & E5       & -4.7 &  15.71 &  10.13 &  10.88 &\nodata &$ <
     7.01 $&  1.49 & 13.2   & \nodata \\
     4486 & Cote'/3649 &$E1^+pec$& -4.3 &  16.07 &  10.86 &  11.44 &\nodata &$ <
     7.03 $&  1.14 & 15.1   &  -13.26 \\
     4552 & Kennicutt/159 & E0-1     & -4.6 &  15.34 &  10.26 &  11.03 &\nodata &$ <
     7.18$&  1.39 &  8.9   &  -14.27 \\
     4564 & Fabbiano/20371 & E        & -4.8 &  15.00 &  9.81  &  10.53 &\nodata &
     \nodata  &  1.52 &  3.6   & \nodata \\
     4621 & Cote'/3649 & E5       & -4.8 &  18.28 &  10.44 &  11.17 &\nodata &
     \nodata  &  1.40 & \nodata& \nodata \\
     4660 & Cote'/3649 & E        & -4.7 &  12.82 &  9.55  &  10.28 &\nodata &
     \nodata  &  1.43 & \nodata& \nodata \\
     5813 & Fazio/69 & E1-2     & -4.8 &  32.21 &  10.73 &  11.21 &\nodata &$ <
     7.64 $&  1.51 & 15.1   &  -14.01 \\
     5831 & Surace/3403 & E3       & -4.8 &  27.16 &  10.18 &  10.84 &\nodata &$ <
     7.79 $&  1.77 &  3.2   &  -15.06 \\
     5845 & Fabbiano/20371 & E:       & -4.8 &  25.94 &  10.31 &  11.15 &\nodata &$ <
     7.45 $&  1.51 &  8.9   & \nodata \\
     5846 & Fazio/69 & E0-1     & -4.7 &  24.89 &  10.73 &  11.36 &\nodata &$ <
     7.72 $&  1.34 &  7.9   &  -13.85 \\
     5982 & Surace/3403 & E3       & -4.8 &  40.18 &  10.59 &  11.30 &  7.42  &$ <
     7.53 $&  1.63 & 13.2   &  -14.79 \\
\\
     0474 & Zezas/20140 &$S0^0(s)$ & -2.0&  33.46 &  10.36 &  10.77 & \nodata &$ <
     7.63$&  1.68 & \nodata&  -14.09 \\
     1023 & Fazio/69 &$SB0^-(rs)$& -2.7 &  11.42 &  10.59 &  10.96 &  9.32  &$ <
     6.69 $&  1.49 &  4.7   &   -14.1 \\
     2685 & Rieke/40936 &$(R)SB0^+pec$& -1.1 &  15.63 &   9.83 &  10.41 &  9.26  &
     7.50  &  2.01 & \nodata&  -13.48 \\
     3032 & Young/20780 &$SAB0^0(r)$ & -1.8 &  21.98 &  9.69  &  10.17 &\nodata &
     8.42  &  4.57 &  0.9   &  -13.79 \\
     3156 & Surace/3403 & S0:      & -2.4 &  22.39 &   9.81 &  10.23 &\nodata &
     7.62  &  3.92 & \nodata&  -13.71 \\
     3384 & Surace/3403 &$SB0^-(s) $ & -2.7 &  11.59 &  9.98  &  10.78 &\nodata &$ <
     7.07$&  1.94 &  3.2   &  -14.67 \\
     3489 & Fazio/69 &$SAB0^+(rs)$ & -1.3 &  12.08 &  9.96  &  10.54 &\nodata &
     7.19  &  2.82 &  1.7   &  -12.95 \\
     4150 & Fazio/69 &$ S0^0(r)$    & -2.1 &  13.74 &  10.30 &  10.02 &  6.40  &
     7.82 &  3.47 &  1.5   &   -13.9 \\
     4382 & Cote'/3649 &$ S0^+(s)pec$ & -1.3 &  18.45 &  10.77 &  11.70 &\nodata &$
     < 7.34$&  2.16 &  1.7   & \nodata \\
     4459 & Cote'/3649 &$ S0^+(r) $   & -1.4 &  16.14 &  10.11 &  10.81 &\nodata &
     8.22 &  2.16 &  1.9   &  -14.07 \\
     4477 & Rieke/40936 & SB0(s):  & -1.9 &  17.06 &  10.19 &  10.87 &\nodata &
     7.55 &  1.65 & \nodata&  -13.54 \\
     4526 & Fazio/69 &$ SAB0^0(s): $  & -1.9 &  16.90 &  10.52 &  11.21 &\nodata
     &    8.60  &  1.86 &  2.8   &  -13.74 \\
     4570 & Cote'/3649 & S0 sp    & -2.0 &  25.90 &  10.37 &  10.73 &\nodata &$
     <7.75 $&  1.45 & \nodata&  -16.02 \\
     7457 & Fazio/30318 &$S0^-(rs)? $ & -2.6 &  13.24 &  9.97  &  10.32 &$< 6.18$&
     6.66  &  2.28 & \nodata&  -14.49 \\
\enddata
\tablenotetext{a}{Hubble type is taken from de Zeeuw et al. (2002). 
$^b$ The morphological type T is taken from the HyperLeda database.
$^c$ Distances are taken from Tonry et al. (2001). $^d$ Luminosities in the $Ks$ 
band have been derived using the flux density $F_{Ks}$ published by the 2MASS survey.
$^e$ The neutral hydrogen masses $M_{HI}$ are taken from 
Morganti et al. (2006).
$^f$ Molecular gas masses $M_{H_2}$ are taken from Combes
  et al (2007).
$^g$ $H\beta$ indexes (estimated within $R_e/8$) from
  Kuntschner et al. (2006).
$^h$ Ages are taken from McDermid et al. (2006).
$^i$ $H\beta$ emission 
fluxes are taken from Sarzi et al. (2006)}.

\end{deluxetable}

\clearpage

\begin{deluxetable}{lrrrrrrrrc}
\tabletypesize{\scriptsize}
%\rotate
%\tablecaption{MIPS Far-infrared photometry }
\tablecaption{}
\tablewidth{0pt}
\tablehead{
\colhead{Name} &\multicolumn{3}{c}{Flux Density} &\colhead{} &
\multicolumn{3}{c}{Log $L_{\lambda}\Delta {\lambda}$} 
& \d{} &  \colhead{SFR\tablenotemark{a}} \\
\cline{2-4}\cline{6-8}\\
\colhead{(NGC)} & 24 $\mu$m & 70 $\mu$m & 160 $\mu$m & & 24 $\mu$m  &
70 $\mu$m  & 160 $\mu$m & & \colhead{(24$\mu m$)}    \vspace{0.1cm}\\
 & (mJy) & (mJy) & (mJy) &  & ($erg~s^{-1}$) &  ($erg~s^{-1}$) &
($erg~s^{-1}$) & & ($M_\odot~yr^{-1}$)
}
\startdata
0821   &     15.1$\pm$4.8 &    $< 15.6$  &    $<24.8$   &     &
40.46&$<40.10$ &$<39.84$&  &  \nodata               \\
%2699 &         $\pm$    &    $\pm$     &    $\pm$     &     &      &
%&     &  &  \nodata               \\
2768  &     46.6$\pm$8.2 &694 $\pm$26   &377 $\pm$29   &     &
40.89&41.68 &40.95&  &  $2.03 \times 10^{-2}$\\
2974  &     62.5$\pm$12.4&682 $\pm$18   &1979$\pm$54   &     &
40.98&41.64 &41.64&  &  \nodata               \\
3377  &     17.3$\pm$5.4 &80.6$\pm$5.7  &70.7$\pm$9.2  &     &39.86
&40.15 &39.63&  &  \nodata               \\
3379  &     65.5$\pm$8.8 &60.5$\pm$8.0  &59.1$\pm$8.2  &     &40.38
&39.97 &39.50&  &  \nodata               \\
3608  &     18.4$\pm$4.2 &66.2$\pm$8.2  &  94$\pm$21   &     &
40.50&40.68 &40.37&  &  $1.18 \times 10^{-2}$ \\
4278  &     43.9$\pm$12.7& 790$\pm$18   &1356$\pm$28   &     &40.57
&41.45 &41.22&  & $1.49 \times 10^{-2}$  \\
4374  &     65.3$\pm$8.6 & 588$\pm$21   & 487$\pm$21   &     &40.86
&41.44 &40.89&  & \nodata                \\
%4387 &         $\pm$    &    $\pm$     &    $\pm$     &     &      &
%&     &  & \nodata                \\
4458  &     3.2 $\pm$1.1 &    $<16.5$   &    $<21.8$   &     &39.50
&$<39.83$ &$<39.49$&  & \nodata                \\
4473  &     25.8$\pm$6.7 &    $<8.0$    &    $<15.1$   &     &40.32
&39.44 &39.25&  &  \nodata               \\
4486  &    151  $\pm$ 9  & 460$\pm$14   &815 $\pm$31   &     &41.11
&41.22 &41.00&  &  $4.03 \times 10^{-2}$ \\
4552  &     57.3$\pm$7.8 &92.1$\pm$10.2 &171 $\pm$16   &     &40.65
&40.48 &40.28&  &  $1.08 \times 10^{-2}$ \\
4564  &     23.5$\pm$3.7     &  \nodata     &  \nodata     &
& 40.24 & \nodata&\nodata&  & \nodata                \\
4621  &     34.2$\pm$6.3 &31.7$\pm$5.7  &43.4$\pm$6.6  &     &40.58
&40.17 &39.84&  &    \nodata             \\
4660  &     15.2$\pm$4.3 &36.4$\pm$6.2  &53.7$\pm$8.2  &     &39.92
&39.92 &39.62&  & $2.93 \times 10^{-3}$  \\
%5198 &         $\pm$    &    $\pm$     &    $\pm$     &     &      &
%&     &  & \nodata                \\
5813  &     14.9$\pm$4.9 &58.4$\pm$7.6  &34.8$\pm$5.9  &     &40.71
&40.93 &40.24&  & \nodata                \\
5831  &     14.5$\pm$4.6 &  $<15.8$     &    $<21.6$   &     &
40.55&$<40.21$ &$<39.88$&  & $1.46 \times 10^{-2}$  \\
%5838 &         $\pm$    &    $\pm$     &    $\pm$     &     &      &
%&     &  &  \nodata               \\
5845  &      7.6$\pm$3.2 & 103$\pm$12   & 154$\pm$13   &     &40.23
&40.98 &40.69&  &  \nodata               \\
5846  &     38.5$\pm$6.2 &102 $\pm$10   &117 $\pm$12   &     &40.90
&40.94 &40.54&  &  $3.65 \times 10^{-3}$ \\
5982  &     13.5$\pm$5.1 &37.2$\pm$7.2  &59.9$\pm$11.7 &     &40.86
&40.92 &40.66&  & $7.17 \times 10^{-3}$  \\
\\
%0454  &         $\pm$    &    $\pm$     &    $\pm$     &     &      &
%&     &  & \nodata                \\
%0455  &         $\pm$    &    $\pm$     &    $\pm$     &     &      &
%&     &  & \nodata                \\
0474   &      4.9$\pm$2.2 &36.2$\pm$7.5  &97.6$\pm$22.6 &     &40.26
&40.75 &40.72&  & \nodata                \\
%0524  &         $\pm$    &    $\pm$     &    $\pm$     &     &      &
%&     &  & \nodata                \\
1023  &     58.9$\pm$5.6 &    $<15.6$   &    $<24.8$   &     &40.40
&$<39.45$&$<39.19$&  & \nodata                \\
%2549 &         $\pm$    &    $\pm$     &    $\pm$     &     &      &
%&     &  &   \nodata              \\
2685  &      \nodata     & \nodata      &  \nodata     &
&\nodata&\nodata&\nodata&  & \nodata                \\
%2695 &         $\pm$    &    $\pm$     &    $\pm$     &     &      &
%&     &  &  \nodata               \\
3032  &    149  $\pm$8   &2640$\pm$34   &2768$\pm$73   &     &41.38
&42.25 &41.80&  &  $1.97 \times 10^{-1}$ \\
3156  &     16.3$\pm$6.2 & 242$\pm$8    & 185$\pm$11   &     &40.43
&41.23 &40.64&  & $2.41 \times 10^{-2}$  \\
3384  &    60.2 $\pm$7.8 & 42.5 $\pm$ 17.2&  \nodata     &     &40.43
& 39.90&\nodata&  & \nodata             \\
%3414 &         $\pm$    &    $\pm$     &    $\pm$     &     &      &
%&     &  &  \nodata               \\
3489  &    89.8 $\pm$6.7 &1673$\pm$18   &2559$\pm$38   &     &40.64
&41.53 &41.25&  & $3.38 \times 10^{-2}$  \\
4150  &    66.9 $\pm$9.5 &1432$\pm$28   &1672$\pm$35   &     &40.62
&41.57 &41.18&  & $4.02 \times 10^{-2}$  \\
%4262 &         $\pm$    &    $\pm$     &    $\pm$     &     &      &
%&     &  &  \nodata               \\
%4270 &         $\pm$    &    $\pm$     &    $\pm$     &     &      &
%&     &  &  \nodata               \\
4382  &    51.1 $\pm$6.7 &22.5$\pm$6.5  &18.7$\pm$9.1  &     &40.76
&40.03 &39.48&  & \nodata                \\
4459  &    105  $\pm$8   &2286$\pm$19   &3147$\pm$43   &     &40.96
&41.92 &41.59&  & $6.69 \times 10^{-2}$  \\
4477  &   12.6$\pm$3.2       & $ <18$    & $<22$      &
& 40.08& $<39.86$ & $<39.48$&  & \nodata            \\
4526  &    262  $\pm$12  &7750$\pm$32   &12460$\pm$67  &     &41.39
&42.49 &42.23&  & $1.68 \times 10^{-1}$  \\
4570  &     18.3$\pm$6.0 &    $<18.3$   &    $<27.6$   &     &40.61
&40.22 &39.94&  & $2.50 \times 10^{-2}$  \\
%5308 &         $\pm$    &    $\pm$     &    $\pm$     &     &      &
%&     &  &  \nodata               \\
%7332 &         $\pm$    &    $\pm$     &    $\pm$     &     &      &
%&     &  &  \nodata               \\
7457  &      8.2$\pm$2.5 &    $<15.9$   &    $<23.5$   &     &39.68
&$<39.59$&$<39.29$&  & \nodata \\

\enddata
\tablenotetext{a}{The mean scatter among the SFRs for elliptical galaxies in 
the upper section of this table, $0.014$ $M_{odot}$ yr$^{-1}$, 
suggests that the (somewhat larger) 
star formation rates for S0 galaxies in the lower section of the 
table are uncertain to 
$\sim \pm 0.02$ $M_{\odot}$ yr$^{-1}$.}
%\tablenotetext{a}{Galaxies are selected from the following Spitzer
%observing programs:
%GO program number 3649, P. Cote' (PI);
%}
%\tablenotetext{b}{The morphological type T is taken from the
%HyperLeda database.}
%\tablenotetext{c}{Distances are calculated with $H_0=70$ km s$^{-1}$
%Mpc$^{-1}$.}

\end{deluxetable}

%\end{document}
\clearpage

%\vskip1.in
\begin{figure}[ht]%1 
%\figurenum{1}
\centering
%\vskip2.in
%%\includegraphics[bb=90 216 522 569,scale=0.9,angle= 270]
%\includegraphics[bb=90 166 522 519,scale=1.0,angle= 0]
%\includegraphics[bb=50 216 422 669,scale=0.8,angle=0]{f1.ps}
\includegraphics[bb=250 216 422 769,scale=1.0,angle=0]{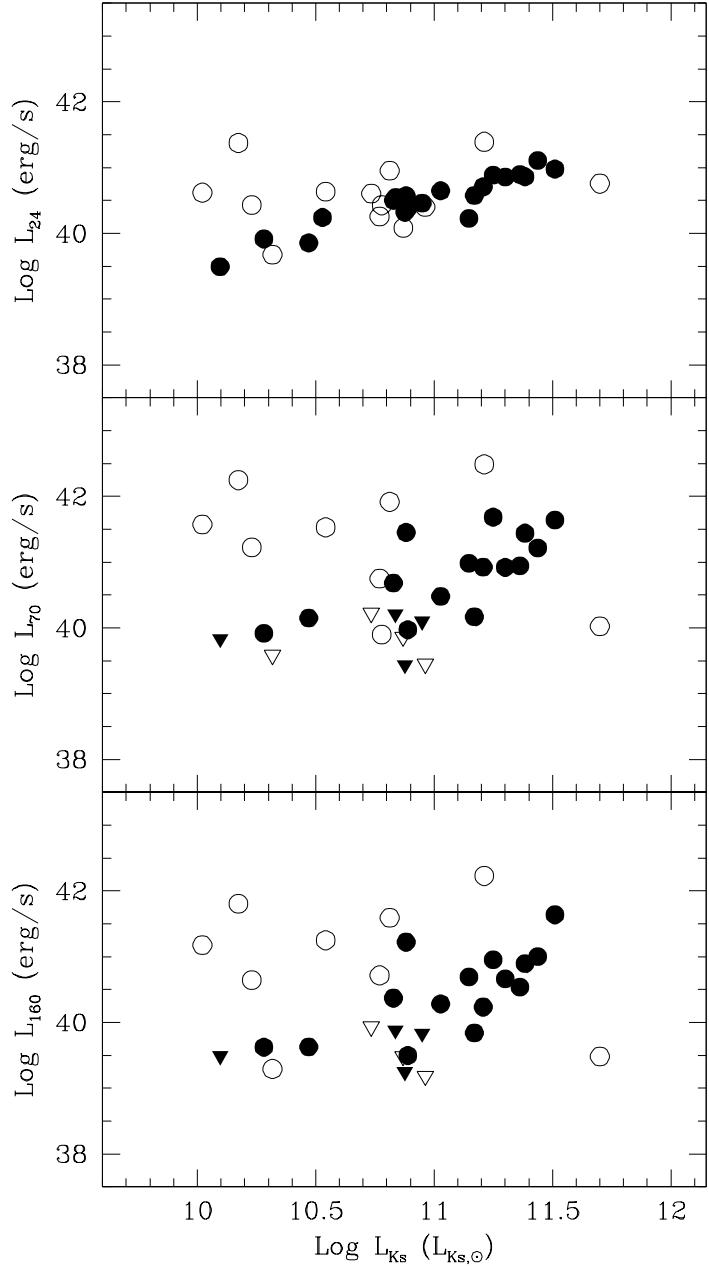}
\vskip.7in
\caption{
SAURON galaxy luminosities in 24, 60 and 170$\mu$m {\it Spitzer} 
bandpasses plotted against 2MASS Ks band luminosities.
Detected E galaxies are filled circles and S0 galaxies are open
circles. 
Triangles represent upper limits.
}
\label{f1}
\end{figure}

\clearpage
\begin{figure}[ht]%1
%\figurenum{2}
\centering
\vskip1.in
\includegraphics[bb=250 136 422 669,scale=1.0,angle=0]{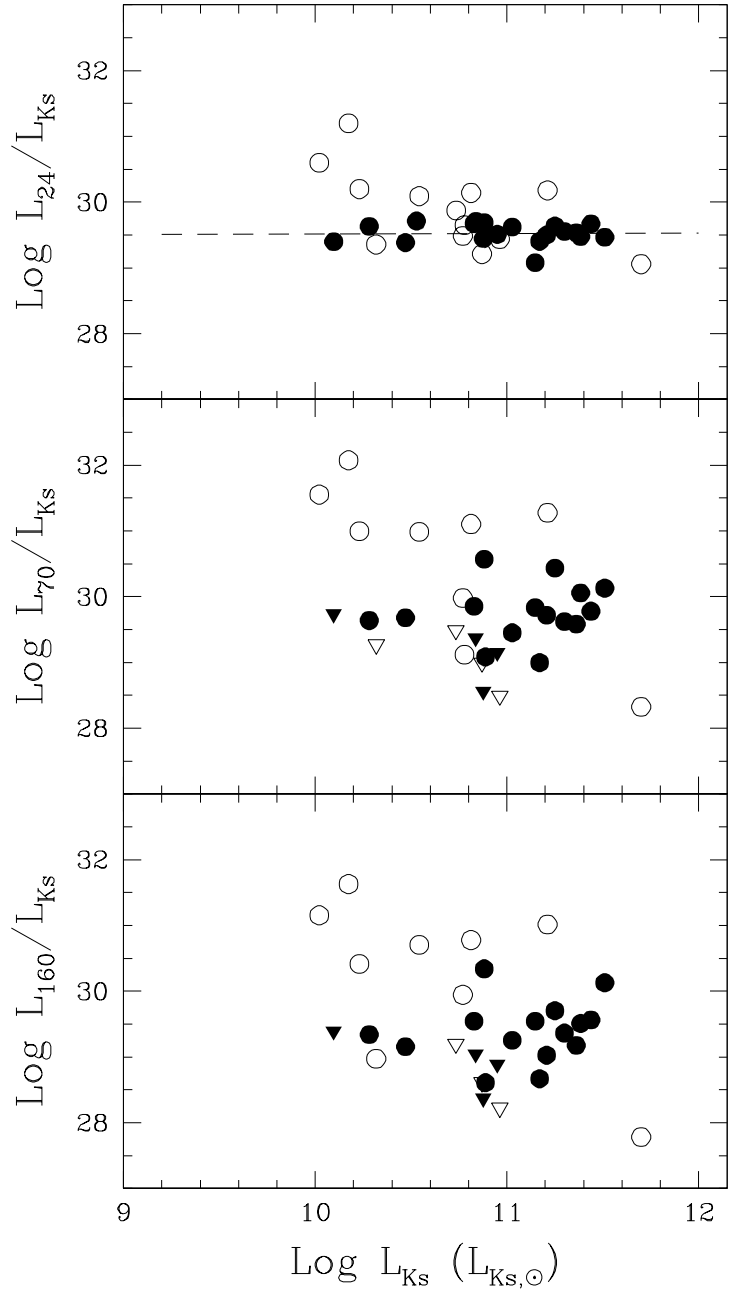}
\vskip.7in
\caption{
Plot similar to Figure 1 but with $L_{FIR}/L_{Ks}$ plotted against 
$L_{Ks}$. The dashed line in the upper panel is a least squares 
fit for just the E galaxies (filled circles). 
The Ks - 24$\mu$m color is nearly constant for old stellar populations 
in E galaxies regardless of $L_{Ks}$.
In general S0 galaxies (open circles) 
have a much wider range of $L_{FIR}$ than E galaxies.
}
\label{f2}
\end{figure}

\clearpage
\vskip2.in
\begin{figure}[h]%1
%\figurenum{3}
\centering
\vskip2.in
\includegraphics[bb=250 216 422 569,scale=1.2,angle=0]{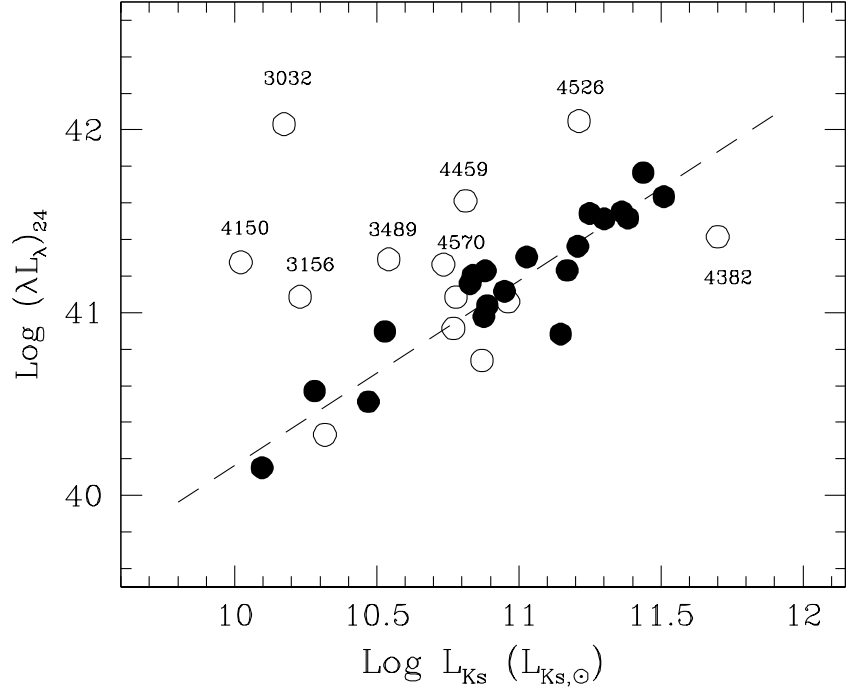}
\vskip0.3in
\caption{
Plot of $(\lambda L_{\lambda})_{24}$ 
against Ks-band luminosity $L_{Ks}$, 
showing a tight correlation among E galaxies and 
uncorrelated 24$\mu$ excesses 
for many S0 galaxies.
}
\label{f3}
\end{figure}

\clearpage
\vskip2.in
\begin{figure}[h]%1
%\figurenum{4}  
\vskip2.in
\includegraphics[bb=120 136 292 569,scale=1.2,angle=0]{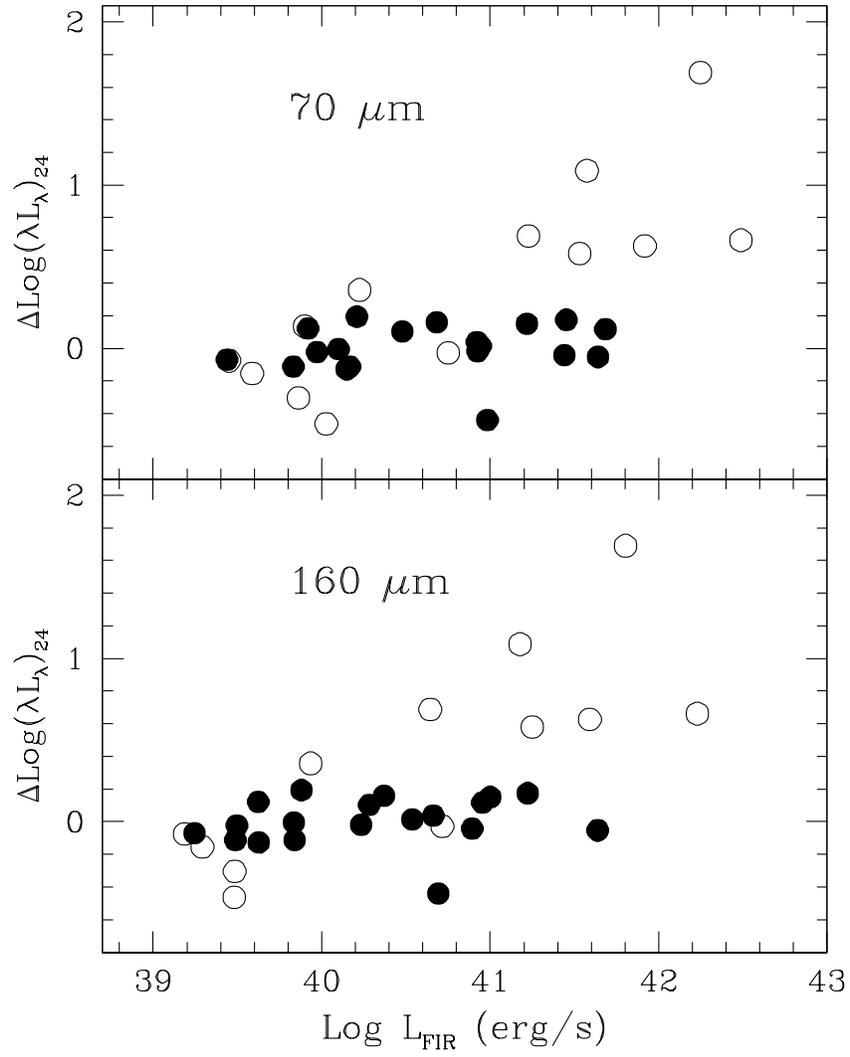}
\vskip0.3in
\caption{
Residuals
$\Delta{\rm Log}(\lambda L_{\lambda})_{24}$
in Figure 3 plotted against passband luminosities 
at 70 and 160$\mu$m. In this and subsequent figures $L_{FIR}$ 
represents either $L_{70}$ or $L_{160}$.
}
\label{f4}
\end{figure}

\clearpage
\vskip2.in
\begin{figure}[h]%1
%\figurenum{5}
\centering
\vskip2.in
\includegraphics[bb=250 136 422 569,scale=1.2,angle=0]{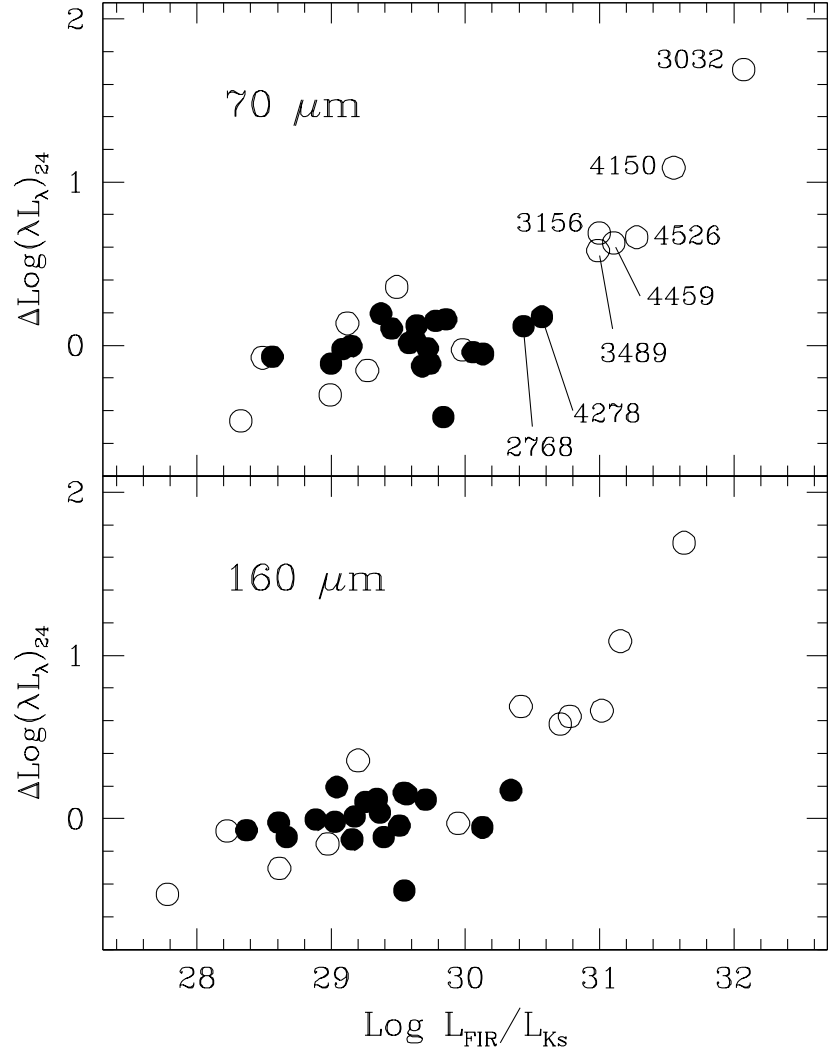}
\vskip0.3in
\caption{
Residuals
$\Delta{\rm Log}(\lambda L_{\lambda})_{24}$
in Figure 3 plotted against passband luminosities
at 70 and 160$\mu$m normalized with $L_{Ks}$. 
Note that the correlation is much tighter than 
in Figure 4.
}
\label{f5}
\end{figure}

\clearpage
\begin{figure}[ht]%1
\centering
\vskip1.in
\includegraphics[bb=250 216 422 669,scale=1.0,angle=0]{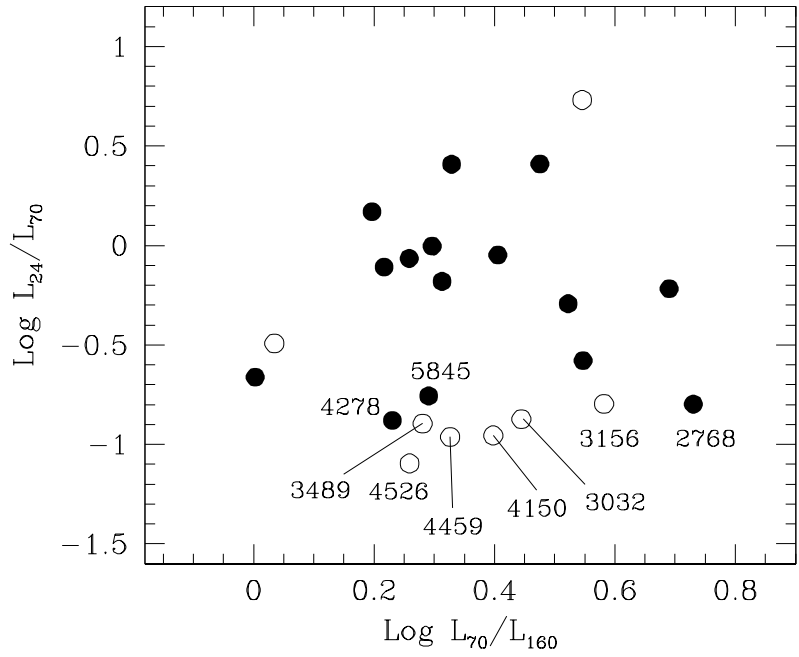}
\vskip.7in
\caption{
Infrared color-color plot for 24, 70 and 160$\mu$m emission 
from Spitzer-SAURON galaxies.
Labeled galaxies with known molecular disks 
lie systematically near the bottom of the plot 
with small Log$(L_{24}/L_{70})$. 
}
\label{f7_01}
\end{figure}

\clearpage
%vskip1.in
\begin{figure}[ht]%1
%\figurenum{7}
\centering
\vskip1.in
\includegraphics[bb=250 136 422 669,scale=1.0,angle=0]{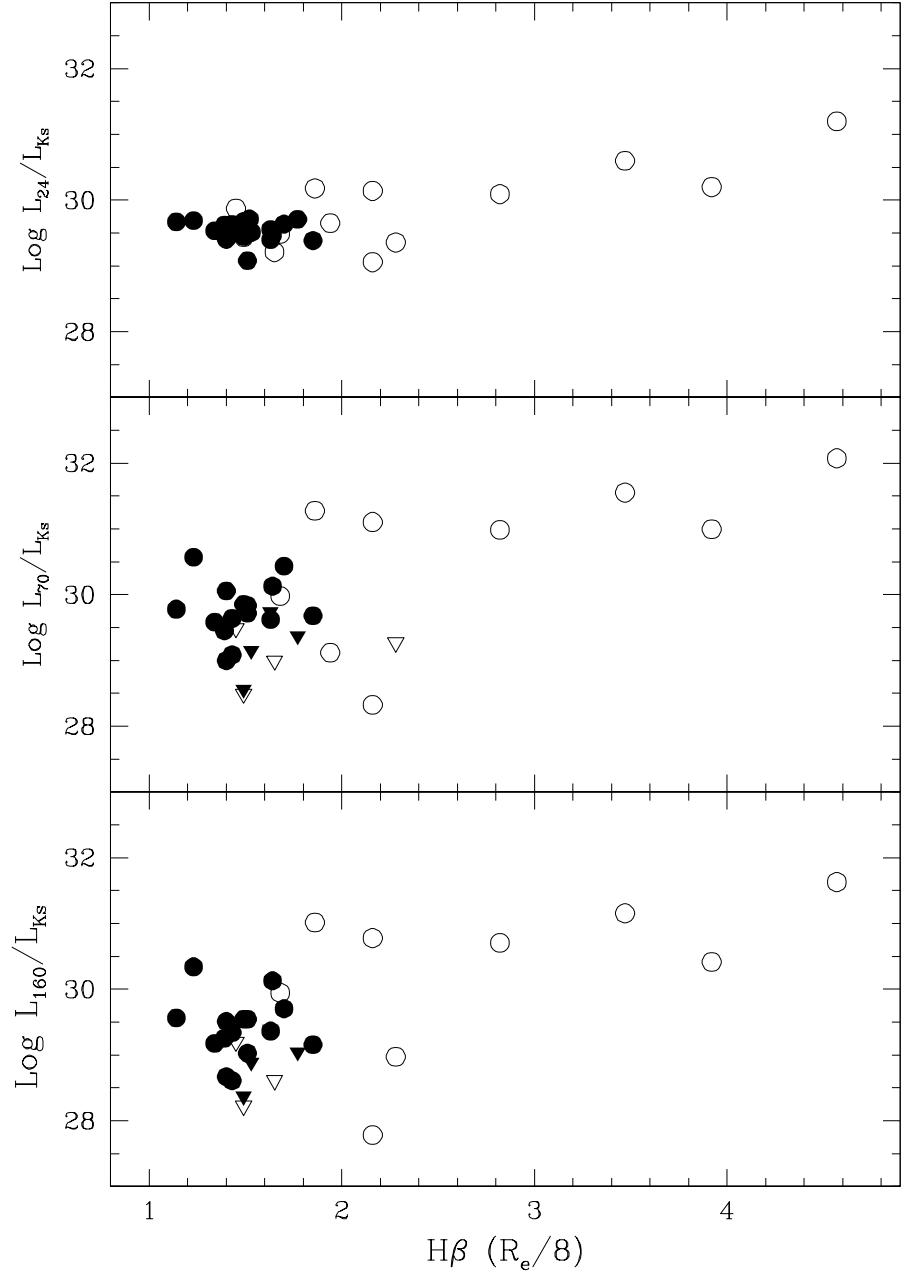}
\vskip.7in
\caption{
Variation of $L_{FIR}/L_{Ks}$ with stellar 
H$\beta$ index (within $R_e/8$).
The Pearson correlation coefficients for the S0 galaxies 
at 24, 70, and 160$\mu$m are $r_{S0} = 0.77$, 0.61 and 0.46 
respectively (discounting upper limits).
}
\label{f7_02}
\end{figure}

\clearpage
\vskip1.in
\begin{figure}[ht]%1
%\figurenum{8}
\centering
\vskip2.in
\includegraphics[bb=250 136 422 600,scale=1.0,angle=0]{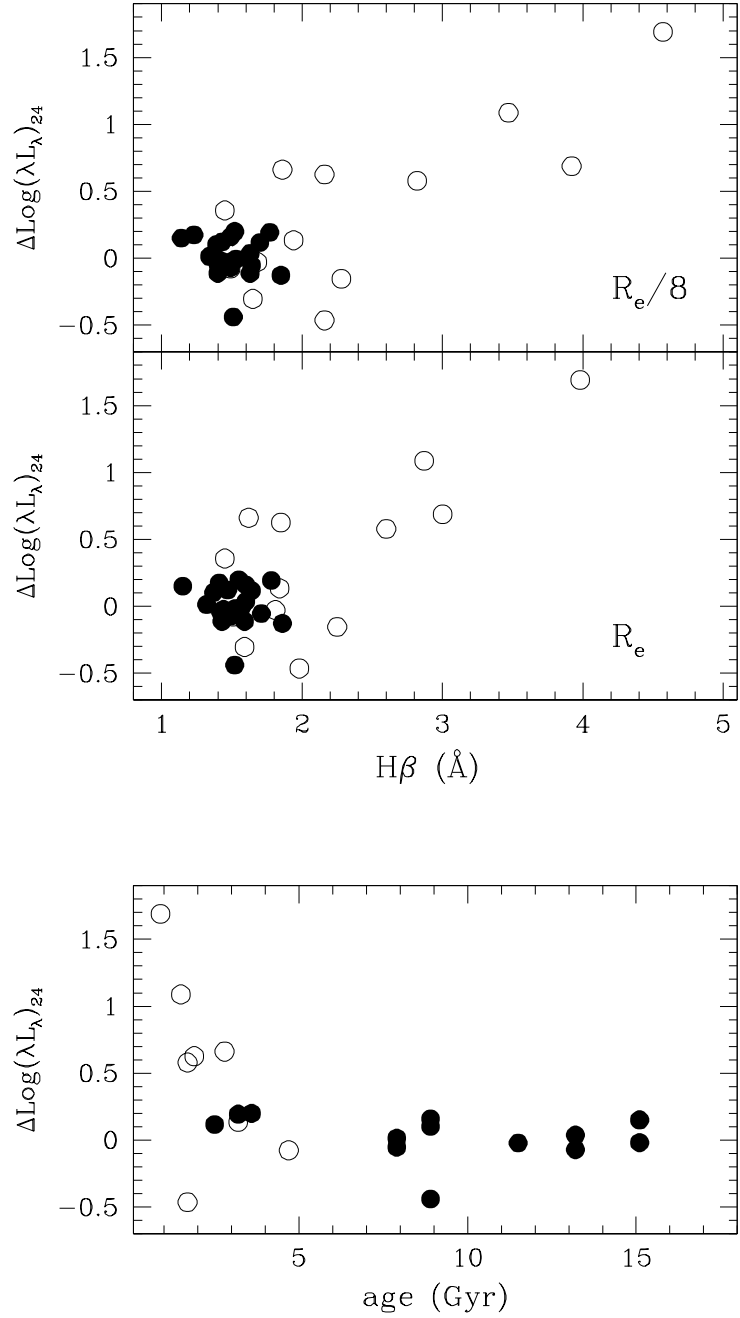}
\vskip.7in
\caption{
{\it Upper panels}: Residuals $\Delta{\rm Log}(\lambda L_{\lambda})_{24}$
in Figure 3 plotted against stellar H$\beta$ index
measured within $R_e/8$ or $R_e$.
Symbols as in Figure 1.
The Pearson correlation coefficients for the S0 galaxies 
are $r_{S0} = 0.77$ for $R_e/8$ and 
$r_{S0} = 0.75$ for $R_e$.
{\it Lower panel}:  Residuals $\Delta{\rm Log}(\lambda
L_{\lambda})_{24}$
in Figure 3 plotted against luminosity-weighted 
stellar age from Table 1.
}
\label{f8}
\end{figure}

\clearpage
\vskip1.in
\begin{figure}[ht]%1
%\figurenum{9}
\centering
\vskip1.in
\includegraphics[bb=250 136 422 669,scale=1.0,angle=0]{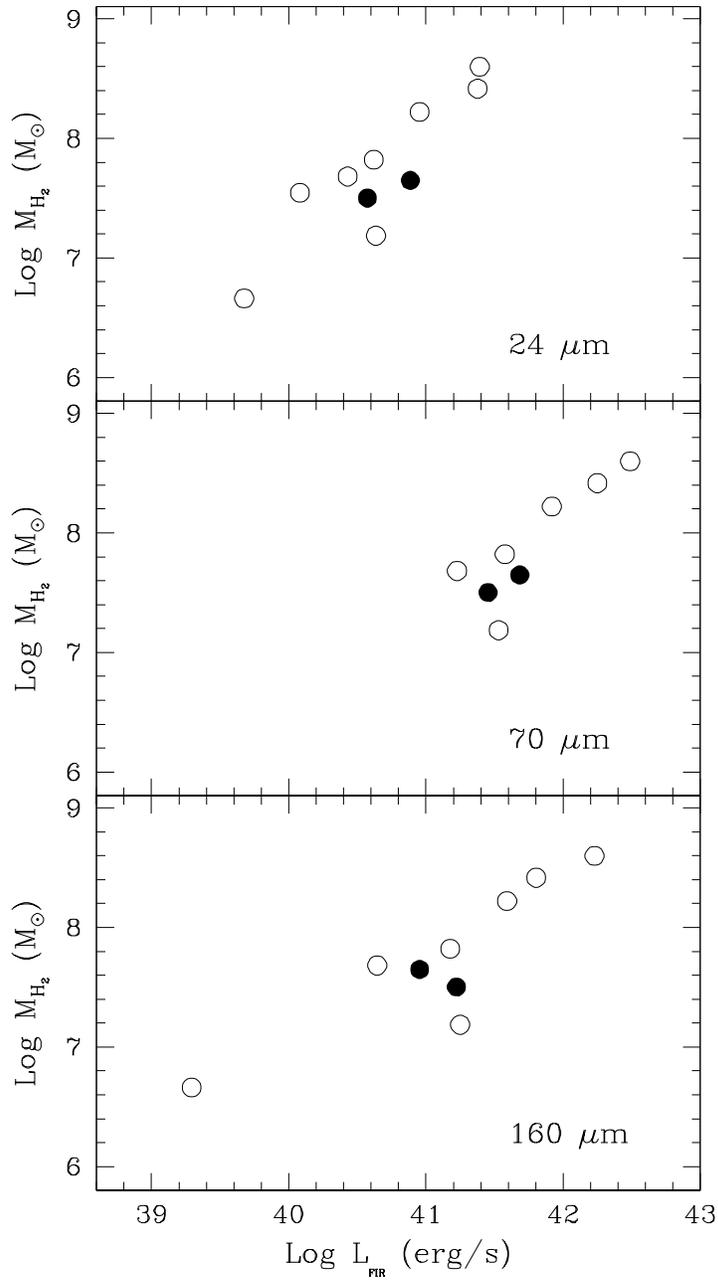}
\vskip.7in
\caption{
Variation of far-IR bandpass luminosities with mass of 
molecular gas. A positive correlation is seen at all three 
wavelengths, 24, 60 and 170$\mu$m.
Symbols as in Figure 1.
}
\label{f9}
\end{figure}

\clearpage
\vskip2.in
\begin{figure}[ht]%1
%\figurenum{10}
\centering
\vskip1.in
\includegraphics[bb=250 136 422 669,scale=1.0,angle=0]{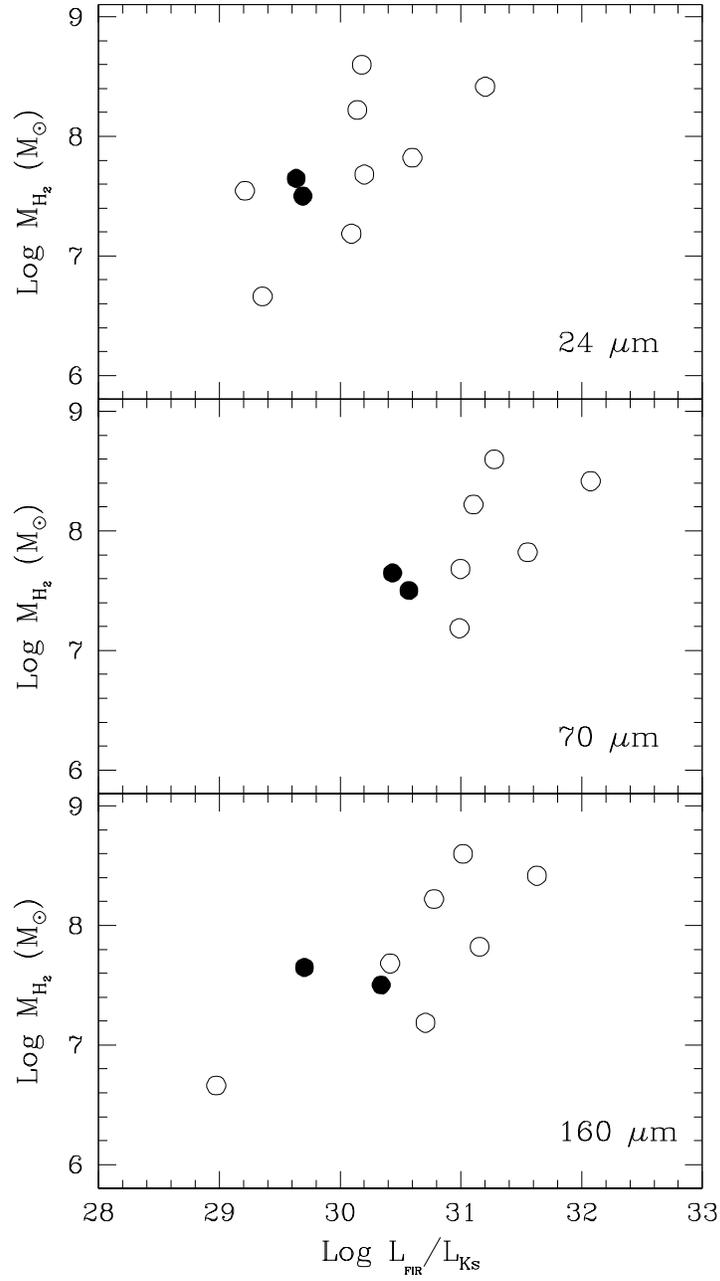}
\vskip1.5in
\caption{
Plots similar to those in Figure 9 but with molecular mass 
plotted against far-IR luminosities normalized with $L_{Ks}$.
Symbols as in Figure 1.
The correlation coefficients for S0 galaxies 
at 24, 70 and 160$\mu$m are 
$r_{S0} = 0.65$, 0.52 and 0.81 respectively.
}
\label{f10}
\end{figure}

\clearpage
\vskip4.in
\begin{figure}[ht]%1
%\figurenum{11}
\centering
\vskip1.in
\includegraphics[bb=250 136 422 669,scale=1.0,angle=0]{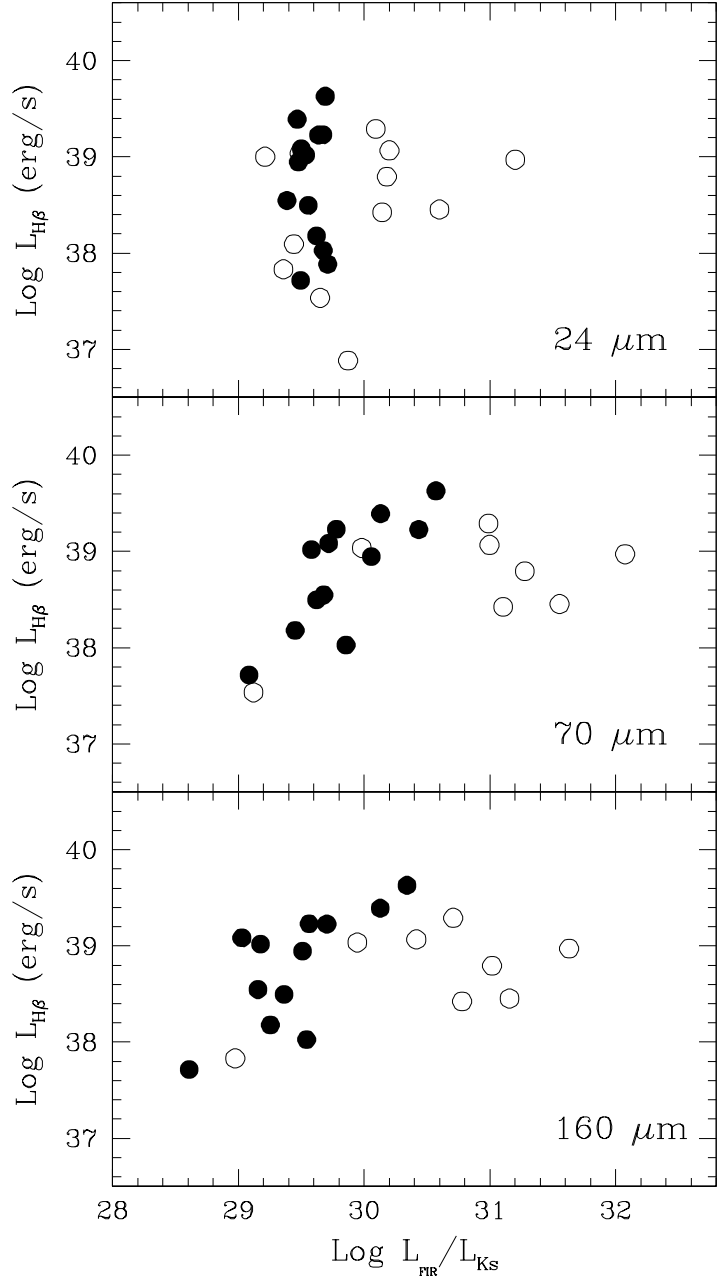}
\vskip.7in
\caption{
H$\beta$ emission line luminosities plotted against 
far-IR band pass luminosities normalized with $L_{Ks}$. 
E galaxies (filled circles) show correlations at 60 and 
170$\mu$m, but S0 data (open circles) show larger scatter 
at all three infrared wavelengths. 
The Pearson correlation coefficients at 24, 70 and 160$\mu$
for S0 and E galaxies are respectively:
$r_{S0} = 0.27$, 0.51 \& 0.32 and 
$r_E = 0.04$, 0.78 and 0.72.
}
\label{f11}
\end{figure}

\clearpage
\vskip4.in
\begin{figure}[ht]%1
%\figurenum{12}
\centering
\vskip1.in
\includegraphics[bb=250 216 422 669,scale=1.0,angle=0]{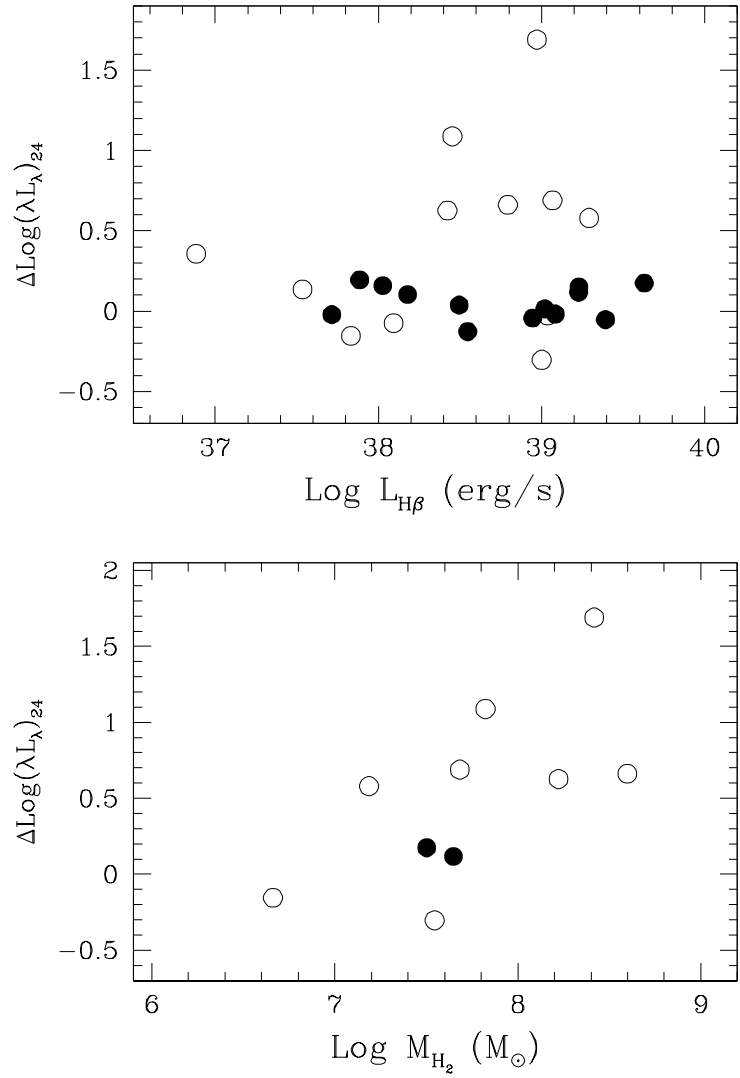}
%\vskip.7in
\caption{
{\it Upper panel:}
Residuals in Figure 3  $\Delta{\rm Log}(\lambda L_{\lambda})_{24}$
plotted against H$\beta$ luminosities. 
E galaxies are filled circles and S0 galaxies are open.
{\it Lower panel:}
Residuals in Figure 3 plotted against mass of molecular gas.
E galaxies are filled circles and S0 galaxies are open.
}
\label{f12}
\end{figure} 

\clearpage
\vskip6.in
\begin{figure}[ht]%1
%%\figurenum{13}
\centering
\vskip4.in
\includegraphics[bb=90 216 422 569,scale=0.9,angle= 270]{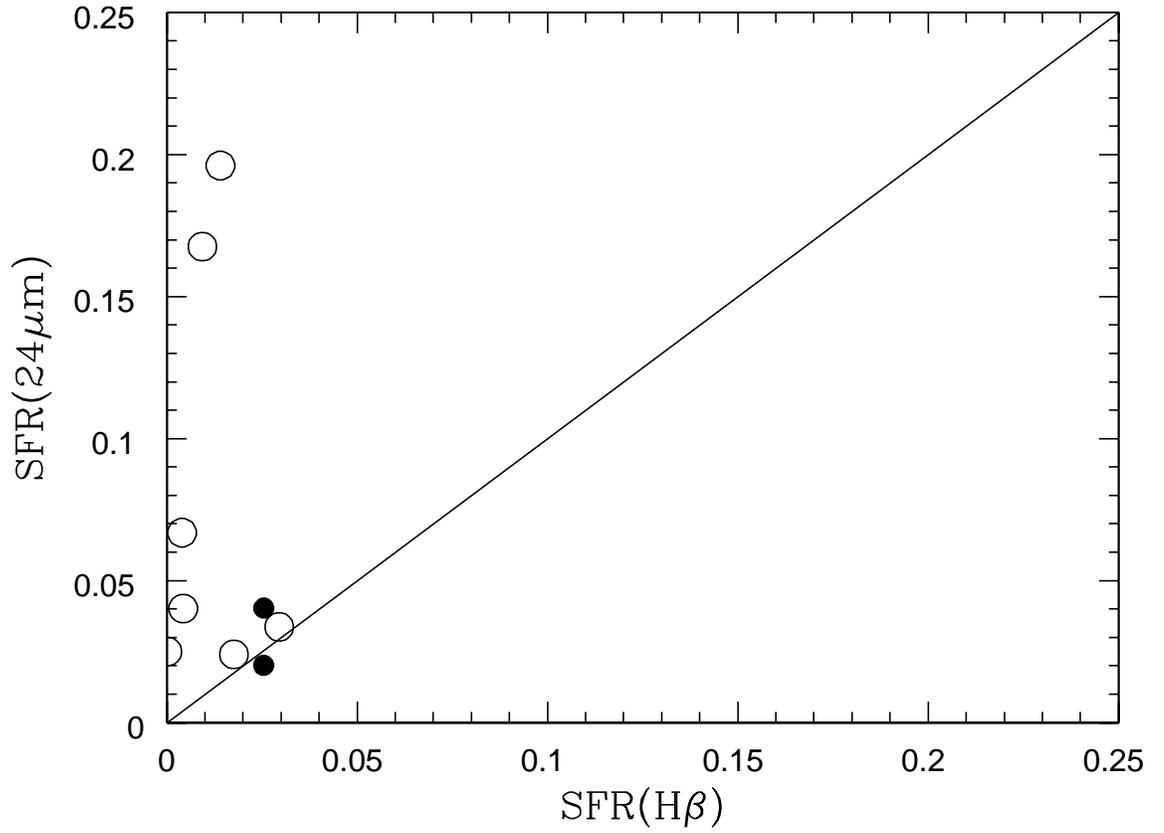}
\vskip.5in 
\caption{
Estimated star formation rates from 24$\mu$m emission 
plotted against star formation rates 
estimated from $H\beta$ emission line luminosities. 
Most S0 galaxies (open circles) have SFR(H$\beta$) $<<$ SFR(24$\mu$m).
Data is available for only 2 E galaxies (filled circles).
}
\label{f13}
\end{figure}

\end{document}